%% file: xi1cma-sub2arch.tex
\documentclass[12pt]{article}

\usepackage{graphics,graphicx}
\usepackage{color}
\usepackage{url}

\usepackage{times}

\newcommand{\xmm}{{XMM-Newton}}

\newcommand{\msim}{\raisebox{-.4ex}{$\stackrel{>}{\scriptstyle \sim}$}}

\newcommand{\xicma}{\mbox{$\xi^1$\,CMa}}

\newcounter{lastnote}

\title{A new type of X-ray pulsar} 

\author
{Lidia M. Oskinova$^{1\ast}$, Yael Naz\'e$^{2}$, Helge Todt$^{1}$,  \\
 David P. Huenemoerder$^3$, Richard Ignace$^4$, \\ 
Swetlana Hubrig$^{5}$,  Wolf-Rainer Hamann$^{1}$ \\
\\
\normalsize{$^{1}$Institute of Physics and Astronomy, University of Potsdam, 
14476 Potsdam, Germany}
\\
\normalsize{$^{2}$FNRS/Univ. of Li\`ege, Dept A.G.O. (GAPHE),
4000-Li\`ege, Belgium}
\\
\normalsize{$^3$ Massachusetts Institute of Technology, Kavli Institute 
for Astrophysics and Space Research,}\\
\normalsize {70 Vassar St., Cambridge, MA 02139, USA}
\\
\normalsize{$^4$ Department of Physics and Astronomy, East Tennessee State 
University,}
\\ 
\normalsize{Johnson City, TN 37663, USA}
\\
\normalsize{$^5$ Leibniz Institute for Astrophysics Potsdam (AIP), An der 
Sternwarte 16}\\
\normalsize{ 14482 Potsdam, Germany}
\\
\normalsize{$^\ast$To whom correspondence should be addressed; 
E-mail: lida@astro.physik.uni-potsdam.de}}

\date{}

\input{bibdefinitions}

\begin{document} 


\baselineskip24pt


\maketitle 

\begin{abstract} 
X-ray emission from stars much more massive than the Sun was discovered
only 35 years ago. Such stars drive fast stellar winds where shocks can
develop, and it is commonly assumed that the X-rays emerge from the
shock-heated plasma. Many massive stars additionally pulsate.  However,
hitherto it was neither theoretically predicted nor observed that these
pulsations would affect their X-ray emission.  Here we report the 
discovery of pulsating X-rays from
the massive B-type star \xicma. This star is a  variable of $\beta$\,Cep
type and has a strong magnetic field.  Our observations with the {\em
XMM-Newton} telescope reveal X-ray pulsations with the same period as
the fundamental stellar pulsation.  This discovery challenges our
understanding of stellar winds from massive stars, their X-ray emission,
and their magnetism.
\end{abstract}

\section{Introduction}
Massive stars $(M_{\rm initial}\,\msim \,8 M_\odot)$ are among the key
players in the cosmic evolution. These hot stars generate most of the
ultraviolet radiation of galaxies  and power their infrared
luminosities.  Massive stars, their winds and their final explosions 
as supernovae provide  significant input of radiative  and mechanical
energy into the  interstellar medium,  inject  
nuclear processed material, and largely drive the evolution of star 
clusters and galaxies.  

The majority of all massive stars are of spectral type B.  Stars with
spectral subtypes  B0-B2 are born with  masses between $8\,M_\odot$ and 
$18\,M_\odot$. They are hot, with effective temperatures $T_{\rm
eff}>15\,000$\,K. While still young and burning hydrogen in their
cores,  these stars oscillate with periods of a few  hours
\cite{st2005}. The physical mechanism that drives these oscillations  is
well understood and is attributed to  changes in the opacity within the
star during the pulsation cycle (``$\kappa$-mechanism'') \cite{dz1993}.
A whole class of such variables  is termed after its prototype 
$\beta$\,Cephei. A B0-B2 type star ends its life
with a supernova explosion, when the stellar  core collapses and
leaves a neutron star as a remnant \cite{heger2003}. Fast rotating,
strongly  magnetic neutron stars may  manifest themselves as radio and
X-ray pulsars.  

All hot massive stars drive stellar winds by their intense radiation.
Photons that are scattered or absorbed in spectral lines transfer their
momentum and thus accelerate the matter to highly supersonic velocities,
typically of the order of 1000\,km\,s$^{-1}$. This driving mechanism is
unstable \cite{lucy1970}. It is generally  believed that the wind
instability may result in wind shocks where  part of the wind material is 
heated to  X-ray emitting temperatures. These wind shocks are often
invoked to explain the X-ray emission which is ubiquitously observed from 
massive stars \cite{feldmeiera1997}. 

Typically, the observed X-ray spectra of single massive stars are
thermal. In high-resolution, X-ray spectra are dominated by slightly
blue-shifted and broad emission lines \cite{wc2007}. These line shapes
can be  explained as originating in a rapidly expanding  
stellar wind, consisting of hot X-ray emitting matter permeated with  
cool matter which attenuates the X-ray field \cite{mac1991,ign2001,osk2006}. 

\begin{figure}
\centering
\includegraphics[width=10cm]{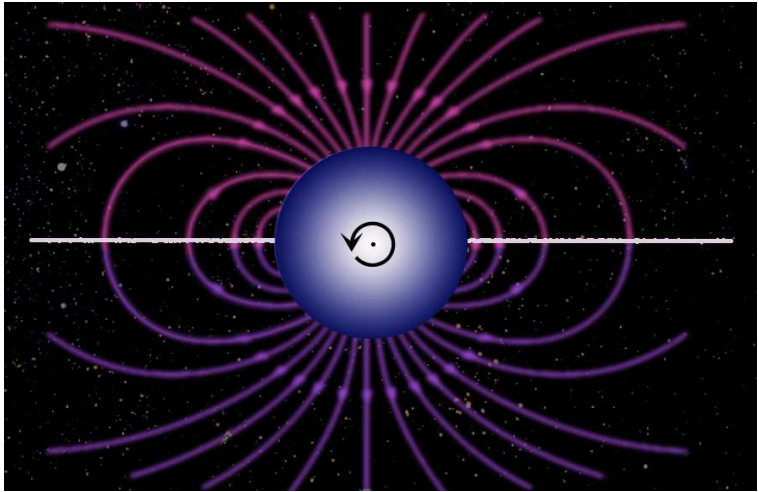}
\caption{Sketch of the rotational and magnetic geometry
for \xicma. From the Earth the star is viewed  nearly rotational 
pole-on (black dot). The star is a magnetic oblique rotator: its magnetic and 
rotational axes are inclined to each other by $\approx 79^\circ$ 
\cite{hub2011}.  Hence the magnetic equator (white line) is always seen 
nearly edge-on. The curved lines illustrate the dipole magnetic field geometry. 
}
\label{fig:dip}
\end{figure}

A small fraction of massive stars, including some B-type stars,  possess
strong, large-scale magnetic fields \cite{hubb2011,Grunhut2012}. According 
to one hypothesis, such stars may have undergone a recent merger process
which powered a dynamo  mechanism  \cite{fer2009,bra2013}. 

Strong magnetic fields  can significantly influence  the dynamics of 
stellar winds \cite{bmb1997}. If the magnetic field has a dipole 
configuration,  the strong field can in effect ``channel''  the  wind
toward the magnetic equator,  where the wind streams from the opposite 
hemispheres collide. The formation of a strong  stationary shock is 
predicted, and it is expected that the X-ray properties 
of stars with strong magnetic fields will be different from those  of 
non-magnetic stars, e.g.\ the  X-ray temperatures for 
magnetically confined winds should be are higher \cite{ud2002}.  

Fluctuations of the X-ray flux  on rotational time-scale are reported for 
single massive stars (magnetic as well 
as non-magnetic) \cite{osk-a2001,naze2010},  and may be related 
to large scale structures in the winds corotating with the star  
\cite{don2001,ign2013}.

\begin{table}
\caption{Parameters of \xicma }
\begin{center}
\begin{tabular}{lr}
\hline \hline
Distance & 424 pc \\
Sp.\ Type & B0.5-B1IV \\
$T_{\rm eff}$ & 27 000 --- 28 000K \\
$\log L/L_\odot$ & 4.5 \\
$R/R_\odot$ & $7\pm 1$ \\
Rotational velocity $v\sin{i}$ & $9 \pm 2$ km\,s$^{-1}$ \\ 
Mass-loss rate $\dot{M}$ & $< 10^{-10}\,M_\odot$\,yr$^{-1}$ \\
Wind velocity $v_\infty$ & $\sim 700$ km\,s$^{-1}$ \\ 
Polar magnetic field strength $B_{\rm p}$ & $5 \pm 1$ kG \\
Pulsation period & 0.2096\,d \\
$L_{\rm X}$ & $3\times 10^{31}$\,erg\,s$^{-1}$ \\
\hline \hline
\end{tabular}
\\
{\small Parameters compiled from the literature, see text for details }
\end{center}
\label{tab:stpar}
\end{table}

The two models briefly outlined above (embedded wind shocks for
non-magnetic objects  and magnetically confined winds)
are widely adopted to interpret the X-ray emission from single hot
massive stars \cite{gn2009}. However, neither of these models  accounts for
the effects of stellar pulsations.

The $\beta$~Cep-type variable \xicma, spectral type B0.5-B1 V-IV
\cite{morel2006}, is an ideal target to study the interactions between
stellar pulsations, magnetic  fields, and stellar wind.  The 
variability of \xicma\ (alias HD\,46328, HR\,2387, HIP\,31125)   is
known for more than 100 years \cite{frost1907}. The period of its radial
velocity variations was established 60 years ago as P=0.2096\,d
\cite{mcnamara1953}. The range of radial velocity is  $\Delta v_{\rm
rad}= 34.2\pm 0.7$\,km\,s$^{-1}$  \cite{shob1973}. \xicma\ is one of the
rare $\beta$\,Cep  variables where the velocity amplitude exceeds the
local speed of sound \cite{saesen2006}.

Together with the radial velocity variations, the star shows periodic
photometric  variability in visual light with a full amplitude of
$\Delta m_{\rm V}=0.034$\,mag  ($\approx$3.2\%) \cite{Wil1954}. Maximum
light occurs about 25\,min after  minimum radius \cite{shob1973}. The
amplitude of photometric variability  increases strongly with
decreasing  wavelength; in the UV near $\lambda 1550$\,\AA\  it amounts
to $0.161$\,mag \cite{lesh1979}.

\xicma\ shows remarkable stability of its pulsational behavior: its
period  is  constant to about 1\,s\,century$^{-1}$ \cite{shob1973}. The
pulsations are non-linear and mono-periodic,  with  only one frequency 
and its first harmonic being  significant 
\cite{hey1994,saesen2006}.

While there is good knowledge of the pulsation properties of \xicma, 
its rotational velocity became established at $v\sin{i} = 10\pm 2$\,km\,s$^{-1}$ 
only recently \cite{morel2006}. This value is consistent  with 
$v\sin{i} = 9\pm 2$\,km\,s$^{-1}$ derived independently from fitting 
high-resolution spectra by means of non-local thermodynamic equilibrium 
(NLTE) models \cite{Lef2010}. 


The fundamental stellar parameters of \xicma\ are typical for its
spectral type B0.5IV \cite{gies1992}. The stellar effective temperature 
changes by $\Delta T_{\rm eff}= 1000$\,K over the pulsation cycle,
ranging  from  27\,000\,K to 28\,000\,K, while $\log{g}$ changes between
3.7 at maximum and 3.8 at minimum
\cite{vanhoof1963,morel2006}. 

The chemical abundances in a sample of $\beta$\,Cep-type variables were
derived from optical spectra using a NLTE line formation code  and line
blanketed LTE  models, but neglecting stellar winds \cite{morel2006}. 
It was found that, similar to some other magnetic early B-type stars,  
nitrogen is overabundant by a factor of 3--4 in \xicma\ as compared to 
the  solar value. 

The advent of UV spectroscopy provided evidence for a stellar wind from
\xicma\ \cite{peters1973}. The study of variability in its UV 
spectra obtained with the IUE observatory did not reveal temporal 
modulations \cite{sch2008}.

The stellar wind parameters of \xicma\ were constrained from the analysis 
of UV spectral lines by means of NLTE  iron-blanketed  model atmospheres
calculated with the PoWR code \cite{hg2004}. The wind from \xicma\ is weak, 
with a mass-loss rate of $\dot{M}<10^{-10}$\,$M_\odot$\,yr$^{-1}$ and a 
terminal velocity of $v_\infty \sim 700$\,km\,s$^{-1}$ \cite{osk2011}. 

The atmosphere models for \xicma\  show that X-rays have a strong effect
on  the ionization of the stellar wind. The observed 
N\,{\sc v} resonance doublet can be
reproduced  only  by  wind models that include X-rays. Interestingly,
the  emission measure (a density squared weighted volume) of the hot 
X-ray emitting plasma in \xicma\ is significantly higher than that of
the  cool gas. Similar conclusions were reached for other late O and
early  B-type stars on the main sequence \cite{hun2012}. 

Despite earlier attempts \cite{rudy1978}, the first firm detection of 
a magnetic field on \xicma\ was achieved only recently \cite{hub2006}. 
Periodic modulations  of the longitudinal magnetic field with a period of
$\approx 2.2$\,d are observed.  This period is identified with the
stellar rotation. Using the known value of $v\sin{i}$
\cite{morel2006,Lef2010}, one can derive the inclination as  $i \approx
3^\circ$. The spectropolarimetric observations are best explained
by assuming that our view of the rotation axis is almost pole-on while
the magnetic axis is inclined by about $79^\circ$ to the rotation axis;
consequently, we have a nearly constant edge-on view of the system's
magnetic equator (see Fig.\,\ref{fig:dip}). The field  has a dipole geometry,
albeit additional  magnetic structures on smaller spatial scales cannot
be  excluded  \cite{hub2011}. A summary of
stellar parameters for \xicma\ is provided in  Table\,\ref{tab:stpar}.
Among pulsating B-type stars with known magnetic field, \xicma\ has  by
far the strongest field with a polar strength of $\sim 5$\,kG \cite{hub2006}. 

While the magnetic field of \xicma\ is strong,  its  stellar wind is
relatively  weak. Therefore, only at distances of more 
than $20\,R_\ast$  the stellar wind can enforce the magnetic 
field lines to  become approximately radial.  Below this distance, 
the stellar wind is controlled by the magnetic  field \cite{osk2011}.

In this work we analyze new, sensitive X-ray observations of \xicma. The
first X-ray survey of $\beta$\,Cep-type stars was performed by  the {\em
Einstein}  observatory \cite{agr1984}.  \xicma\ was detected as the most
X-ray luminous  star in this survey \cite{agr1984,  Grillo1992}. First
X-ray spectra  were obtained with the {\em Rosat} observatory. It was
found that the X-ray spectrum of \xicma\ is somewhat harder than   of
other B-type  stars  \cite{cas1994}. Interestingly, during a 0.56\,h 
exposure with {\em Rosat}, variability at the level of 25\%\ was
noticed, but attributed   to spacecraft wobble \cite{cas1994}.

\section{Results}

The X-ray data for \xicma\ discussed in this paper were taken with the
X-Ray  Multi-Mirror Satellite \xmm\ of the European Space Agency ESA.
Its three telescopes illuminate five different instruments which always
operate simultaneously and independently: RGS1 and RGS2 are Reflection
Grating Spectrometers \cite{rgs2001}, achieving a spectral resolution of
$\sim 0.07$\,\AA\ in the wavelength range 5\,\AA\,--\,38\,\AA.  The
other three focal instruments, forming together  the EPIC camera, are
called MOS1  and MOS2 (Metal-Oxide Semiconductor) and PN (pn-CCDs).
Compared to the RGSs, the EPIC instruments  have a broader wavelength
coverage of 1.2\,\AA\,--\,60\,\AA,  but their spectral resolution is
much lower  ($E/\Delta E\approx 20 - 50$) \cite{mos2001,pn2001}. 

\begin{figure}
\centering
\includegraphics[width=9cm]{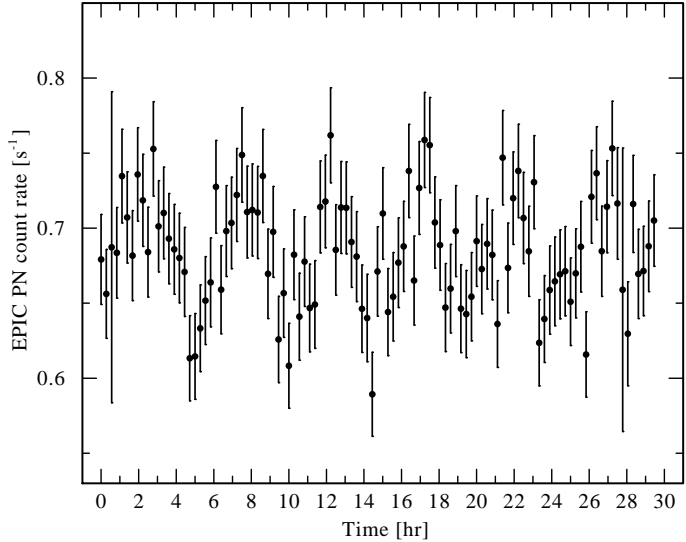}
\caption{
X-ray light curve of \xicma\ in the 0.2\,keV -- 10.0\,keV 
(1.24\,\AA\,--\,62\,\AA) energy band where the   background was
subtracted. The horizontal axis denotes  the time after the beginning of
the observation in hours. The data were binned to 1000\,s. 
The vertical axis shows the count rate as measured by the EPIC PN
camera. The  error bars (1$\sigma$) correspond to the combination 
of the error in the source counts and the background counts. }
\label{fig:lcpn}
\end{figure}

The first dedicated \xmm\ observation of \xicma\ started on  2009-09-03
with an exposure time of 2\,h, which  was far too short  to reveal the
periodic variability \cite{osk2011}. In October 2012, \xmm\ 
continuously observed \xicma\ during nearly 29 hours.  The light curves
and spectra were recorded  by all instruments. The sensitivity of the PN
camera is  superior, and therefore its data are best suited for time
variability studies.

The data reduction involved standard procedures of the \xmm\ Science
Analysis System (SAS) v.13.0.1. The event lists have been filtered, and
time intervals affected by high background were excluded. The useful
exposure time was about 29\,h. The object of our study is sufficiently
isolated,  with no nearby bright optical or X-ray sources. 

The visual magnitude  of \xicma\ is $m_{\rm V}=4.33$\,mag with a
pulsation amplitude of 0.034\,mag  \cite{st2005}. Because of the
brightness of the target in the visual,   the  optical light  blocking
filter of \xmm\ was used  in its ``thick'' mode for our  observations.  
We carefully investigated the potential contamination  of the signal  by
optical/UV light. For the \xmm\ detectors operating  in full-frame mode 
and with the thick filter, no optical loading is  expected for stars
with   $m_{\rm V}>0$\,mag (\xmm\ User Handbook). Moreover, the optical
loading would affect the softest part of the X-ray  spectrum only, but
we don't see different pulsational behavior between softer and harder
parts. Also, while optical pulsations have an amplitude of about 3\%,
the  amplitude of X-ray pulsations is $10\%$. From these considerations,
a  contamination of the signal by optical/UV light can be ruled out.

\subsection{X-ray light curve and comparison with optical light curve}

Visual inspection of the X-ray light curve already reveals the  periodic
variability (see Fig.\,\ref{fig:lcpn}). To quantify this variability, a
statistical analysis is performed.  To obtain a  characterization of
the variability, the light curves in   three  energy bands  (``total'':
0.2\,keV--10.0\,keV (1.24\,\AA\,--\,62\,\AA);   ``soft'':
0.2\,keV--1.0\,keV (12.4\,\AA\,--\,62\,\AA); and ``hard'': 
1.0\,keV--10.0\,keV (1.24\,\AA\,--12.4\,\AA)) are extracted from the
event lists  of all three EPIC instruments separately. This is done
using the {\sc sas}'s  task {\sc epiclccorr}, which provides equivalent
on-axis, full point  spread function count rates  with background
correction. The data are binned in temporal bins of different  duration
(100\,s, 500\,s, 1000\,s, 3600\,s). 

For the further analysis, we  first employ a $\chi^2$ test for several
hypotheses such as constancy, linear variation,  and  quadratic
variation \cite{naze2013}. The PN and MOS1 data in soft and total  bands
display significant variations for all time binnings -- the  probability
that the observed curve occurs by chance is $< 0.01$. Second, an
autocorrelation test is performed to search for recurrence. Variability
on a timescale  of about five hours is detected in PN and  MOS1, but is
not obvious in the MOS2 data due to its poorer sensitivity.  

Finally, a Fourier algorithm is applied \cite{heck1985,gosset2001},
which detects a  period of $4.87\pm 0.09$\,h in the PN data.  This is
much shorter than the plausible rotation period of the star,  but is
conspicuously similar to the stellar pulsation period as derived  from
optical photometry (5.03\,h).     

\begin{figure}
\centering
\includegraphics[width=11cm]{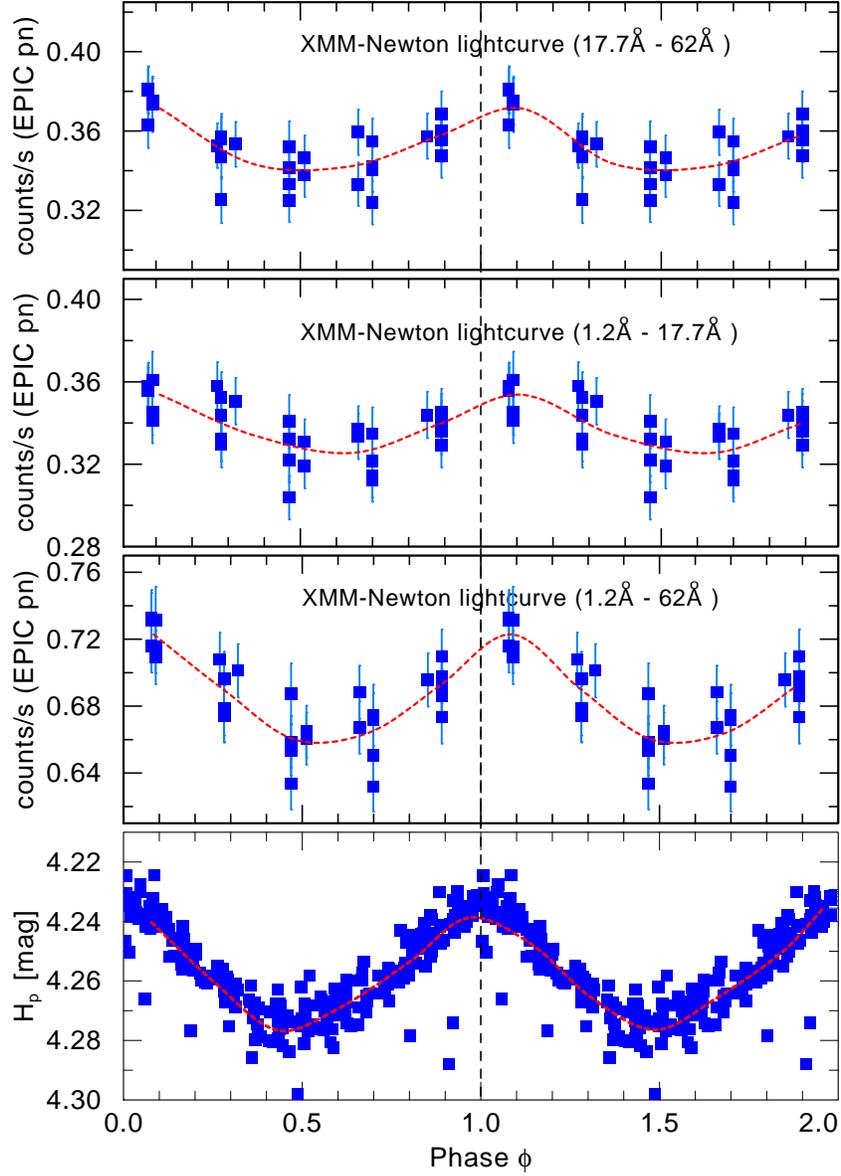}
\caption{X-ray (three upper panels) and optical (lowest panel)
light curves  of \xicma, phased with the stellar pulsation period. The
X-ray  light curve is produced from the data obtained with the  {\em
XMM-Newton} EPIC PN camera, using 1\,h binning.  The X-ray data are
background  subtracted, and the error bars ($1\sigma$) reflect the
combination of the statistical errors in the source counts and the 
background.  The dashed red line interpolates  the averages in phase
bins of  $\Delta\phi=0.1$.  The lower panel shows the {\em Hipparcos}
Catalogue Epoch Photometry data. The abscissa is the magnitude $H_{\rm
p}$ in the {\em Hipparcos}  photometric system (330-900\,nm with maximum at
$\approx 420$\,nm).  The dashed red line interpolates  the averages. }  
\label{fig:pnhip}
\end{figure}

Simultaneous optical and X-ray observations of \xicma\ are not
available. In order to investigate the phase correlation between X-ray
and optical pulsations, we searched for optical photometric data in the
archives. The best and most recent photometry of \xicma\  was obtained
by the space telescope {\em Hipparcos} during the years 1990 - 1993. 
The {\em Hipparcos} photometry was carried out in a wide pass-band,  referred
to as $H_{\rm p}$ \cite{vanleeuwen1997}.  All $H_{\rm p}$ photometric
data were analyzed in a uniform and self-consistent  manner and yielded
the parameters of variability which are compiled in the ``{\em Hipparcos} 
Catalogue Epoch Photometry Data''. The median magnitude of \xicma\ is 
$H_{\rm p}=4.2586$\,mag, and the full variability amplitude is
$0.0398$\,mag. Based on the 5482 observed  cycles, the derived period
is  $P=0.209577\pm 0.000001$\,d and the epoch of  zero phase is  JD(TT)
2448500.0280. We retrieved the {\em Hipparcos} photometric  data for \xicma\
and folded them according to the {\em Hipparcos} ephemeris. The result is
shown in Figure\,\ref{fig:pnhip}  (lower panel).  Comparing this
ephemeris to older measurements from 1954 \cite{Wil1954}, the light
curve is found to be still in phase, i.e.\ the period derived from {\em
Hipparcos} photometry  is very precise and the pulsations have been
stable  over 40  years.  

Our {\em XMM-Newton} data were obtained about 34000 pulsation cycles
after the  {\em Hipparcos}  measurements. However, given the stability
of the \xicma\  light curve \cite{shob1973}, the {\em Hipparcos}
ephemeris can be meaningfully applied for the epoch of our X-ray
observations.  The X-ray light curves were phased with the {\em
Hipparcos}  ephemeris and compared to the optical ones.  
Figure\,\ref{fig:pnhip}   shows  the comparison between X-ray light
curves in full, hard, and soft  bands and the {\em Hipparcos} optical
light curve.   The X-ray  light curve in broad band as well as in softer
and  harder bands turned out to be in phase with the optical within  the
precision  of the measurement and the formal error margin of the
ephemeris  (see Fig.\,\ref{fig:pnhip}).

Typically, the optical light curve of $\beta$\,Cep-type variables  is
similar in shape to the radial-velocity curve but lags in phase. The 
maximum brightness occurs shortly after the stellar radius goes through
its minimum. The results of our analysis show that the same is true for
the X-ray light -- the maximum X-ray brightness is observed close to the
phase when the stellar radius is at its minimum. 

\subsection{Analysis of the low-resolution X-ray spectra}

\begin{figure}
\centering
\includegraphics[width=7cm]{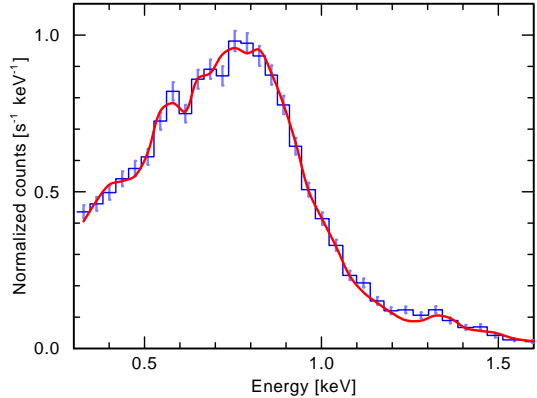}
\caption{ \xmm\ PN spectrum of \xicma\ at phases close to the maximum 
($\phi = 0\pm 0.15$). The observations are represented by the blue
histogram with error bars corresponding to $3 \sigma$. The red curve
shows the best fit model of a three-temperature plasma (see 
Table\,\ref{tab:xspecmod} for the parameters).}
\label{fig:xispmax}
\end{figure}

To define the physical properties of the X-ray emitting plasma in
\xicma, we analyze its X-ray spectra.   As a first step, the low
resolution data recorded by the EPIC cameras  \cite{mos2001,pn2001} are
considered. While individual lines, in general, cannot be resolved with 
the EPIC, the low-resolution spectra cover a broad energy band and allow
to constrain the  temperature and the emission measure of the hot
plasma. For a phase-resolved analysis, the spectra are extracted for
time intervals   close to the pulsation maximum (see
Fig.\,\ref{fig:xispmax}) and the  pulsation  minimum, respectively.  

%

The spectra are analyzed with the standard spectral fitting software
{\sc xspec} \cite{arnaud1996}. In order to reduce the number of free
model  parameters, the interstellar (ISM) neutral hydrogen column
density  is fixed at  $N_{\rm H}=1.4\times 10^{20}$\,cm$^{-2}$ 
according to the interstellar reddening which we obtained from fitting 
the observed spectral energy distribution from the UV to the optical 
\cite{osk2011}, and  in agreement with other works \cite{gud2012}.  The
fit of the X-ray spectra which we obtain with this hydrogen column 
density indicates that no additional absorption intrinsic to the source
is required.    

The spectra were investigated for the presence of an emission  
component described by a power law. Such emission  is the dominant 
component in spectra of X-ray pulsars containing a neutron star. No 
convincing evidence for non-thermal radiation  was  found.  Therefore, 
we adopt a model of thermal plasma in collisional ionization
equilibrium as implemented  in the {\sc apec}  code \cite{apec2001}. To
restrict  the number of free parameters, we  use  only three
temperature  components.  Albeit such a model provides only a simplified
description of the plasma, it can give an insight in the characteristic
temperatures and emission measures.

\begin{table}
\caption{X-ray spectral properties of \xicma\ from fitting the 
low-resolution EPIC spectra at phases close to minimum and maximum}
\begin{center}
\begin{tabular}{lll} \hline\hline
                              & minimum &  maximum \\
\hline 
$kT_1$ [keV]                   & $0.110\pm 0.005$ & $0.115\pm 0.006$ \\ 
EM$_1$ $[10^{53}$\, cm$^{-3}$] & $9.66\pm 1.66$   & $8.48 \pm 1.38$ \\ \hline
$kT_2$ [keV]                   & $0.32\pm 0.02$   & $0.33 \pm 0.02$ \\ 
EM$_2$ $[10^{53}$\,cm$^{-3}$]  & $6.37\pm 0.91$   & $5.71 \pm 0.79$ \\ \hline
$kT_3$ [keV]                   & $0.80\pm 0.02$   & $0.77 \pm 0.02$ \\
EM$_3$ $[10^{53}$\,cm$^{-3}$]  & $3.96\pm 0.30$   & $4.62 \pm 0.30$ \\ \hline
\hline
$\left< kT \right>\equiv \sum_i kT_i\cdot {\rm EM}_i / \sum_i {\rm 
EM}_i$\,[keV] & 0.31 & 0.34 \\
$F_{\rm X}$ [erg\,s$^{-1}$\,cm$^2$] & $1.04\times 10^{-12}$ & $1.15\times 
10^{-12}$ \\
\hline \hline    
\end{tabular}
\end{center}
\label{tab:xspecmod}
\end{table}

The  abundances were set to solar values \cite{aspl2009} for  all
elements included in the {\sc apec} model, except for nitrogen and 
oxygen, which were free fitting parameters.  The best fit resulted in  a
nitrogen overabundance by a factor of  $4.3\pm 0.8$  relative to solar, 
and an oxygen overabundance by a factor of $1.6\pm 0.3$. These abundances
are in agreement with  those  found from the analysis of optical spectra
of \xicma\ \cite{mor2008}. 

The best fit to the observed spectra with a three-temperature plasma
model  is obtained with temperatures ranging between 1.2 and 9 million
Kelvin  (see  Table\,\ref{tab:xspecmod}). Spectral fits indicate that at
the maximum  of the pulsation  cycle the plasma's mean temperature
(i.e.\ the average  temperature weighted with emission measure) is
somewhat higher (by 500\,000\,K) than at minimum.  During the pulsation
cycle, the X-ray flux changes by  $\approx 10\%$.  From the time between
maximum  and  minimum of the light curve ( $2.5$\,h) one can estimate
the cooling rate and use it to constrain the density of X-ray emitting
plasma.   We computed cooling functions  with the  {\sc apec} code, and
found that the observed cooling rate can only  be achieved if the
electron density is higher than  $\approx 3 \times 10^8$\,cm$^{-3}$.

According to our PoWR models for the cool wind of \xicma, such high electron 
densities are encountered only close to the photospheric radius $R_\ast$, and not 
farther than $1.05\,R_\ast$. Assuming that the electron densities  of the hot 
X-ray emitting plasma are not higher than the densities of the ambient cool wind,  
the pulsed X-ray emission should originate from regions very close to the  
stellar photosphere.

\subsection{Analysis of the high-resolution X-ray spectrum}

\begin{figure}
\centering
\includegraphics[width=9cm]{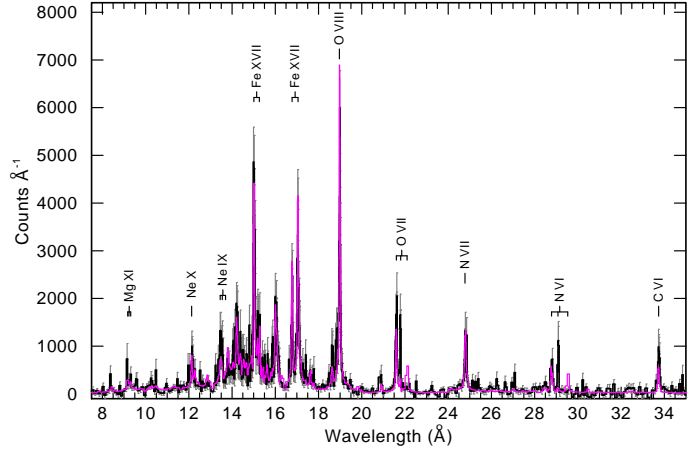}
\caption{Combined RGS1+2 spectrum of \xicma\ (black line) with the 
best fit model (red line). The spectrum is integrated over the full 
exposure time (25.9\,h). Strong emission lines are identified.}
\label{fig:rgscomb}
\end{figure}

The high-resolution X-ray spectra of \xicma\  obtained with the RGS1
and  RGS2 spectrographs \cite{rgs2001}  are
dominated by strong emission lines of metals (see Fig.\,\ref{fig:rgscomb}). 

\begin{figure}
\centering
\includegraphics[width=0.99\columnwidth]{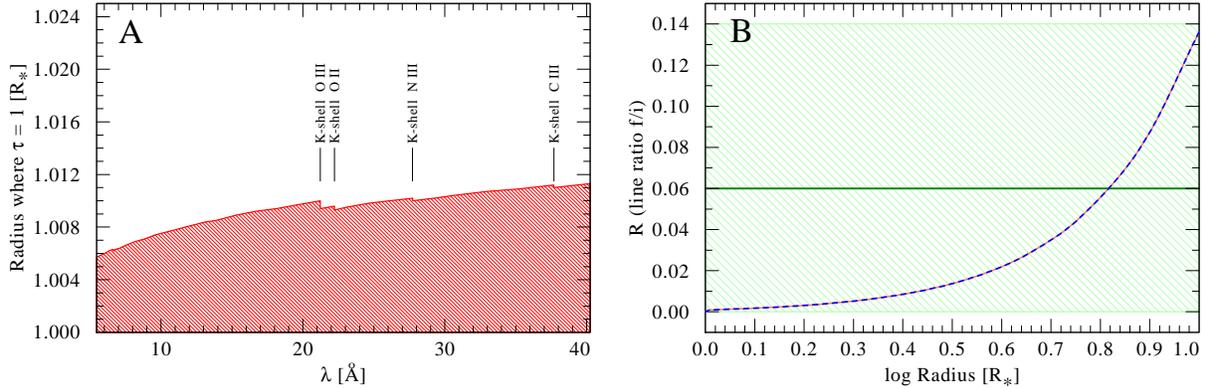}
\caption{{\em Left panel:} Radius (in units of the stellar radius $R_\ast$) 
where the cool wind from \xicma\ becomes optically thin 
in dependence on the wavelength. The calculations were performed
with the PoWR stellar atmosphere code (see text) with stellar 
parameters from Table 1. The prominent photoionization edges 
are identified. {\em Right panel:} Dependence of the line ratio $R=f/i$  
for the O\,{\sc vii} lines as function of the radial location  of the 
emitting plasma. Based on the PoWR model, the red  curve accounts only for 
radiative de-population, while the almost identical blue  curve also  
includes collisional processes under the assumption that the shocked plasma has 
the same density as the cool wind. The measured value 
is shown as horizontal green line, with the green hatched area
representing the $1\sigma$ confidence band of the  measurement. }
\label{fig:tau1}
\end{figure}

X-ray line profiles formed in a rapidly expanding stellar wind would be 
Doppler broadened. These lines can appear characteristically
blue-shifted  and asymmetric, especially when there is significant
absorption of the X-rays in the cool  stellar wind 
\cite{mac1991,ign2001}. To estimate the effects of wind absorption,  
we  computed the radii at which the wind becomes optically thin for
X-rays.    As can be seen from Fig.\,\ref{fig:tau1}, the wind of \xicma\
is optically   thin for X-rays above $1.01\,R_\ast$. Hence no 
significant effects of wind absorption are expected. 

To analyze the line shape, both RGS spectra were combined   and line
moments were calculated. Because of high signal-to-noise ratio, the 
best-suited line for a detailed analysis is O\,{\sc viii} Ly$\alpha$.  
This line is  marginally  broader  than the 
instrumental response. The maximum wind velocity in \xicma\ is  
$\approx 700$\,km\,s$^{-1}$. If X-ray lines were formed in a plasma 
moving with such velocity, the corresponding Doppler line broadening 
would be detectable in the high-resolution spectrum. Only for the
O\,{\sc viii}  Ly$\alpha$ line  a marginal blue-shift is detected, while
the other line are inconclusive.  Hence the emitting plasma is not
moving rapidly, and the attenuation of the X-rays by the cool wind is
low.

The spectrum obtained over the total exposure time and thus covering 
different pulsation phases can be well modeled by a thermal, optically 
thin plasma with temperatures that are consistent with those found from 
fitting the low-resolution spectra as described above (see
Fig.\,\ref{fig:rgscomb}).

\begin{figure}
\centering
\includegraphics[width=8cm]{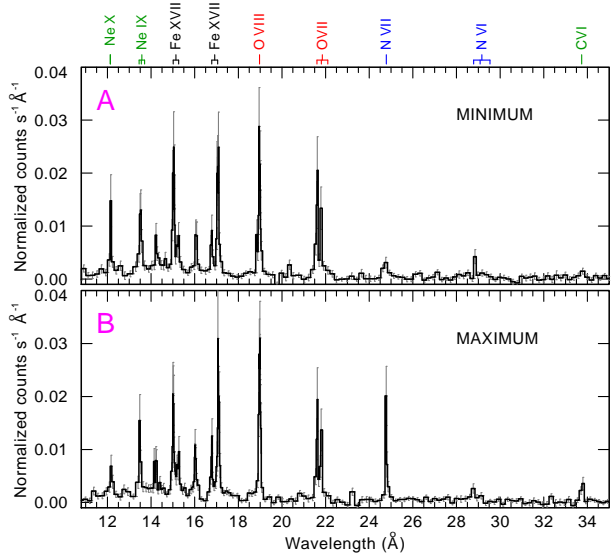}
\caption{Combined high-resolution (RGS1+RGS2) X-ray spectra of 
\xicma\ in phases close to the pulsation minimum (upper panel, 
$\phi = 0.5 \pm 0.17$) and pulsation maximum (lower panel, 
$\phi = 0\pm 0.15$), respectively.  Strong emission lines are 
identified. The error bars correspond to 3$\sigma$.}
\label{fig:rgsphase}
\end{figure}

As a next step, we considered how the high-resolution spectrum varies
between different pulsation  phases (see Fig.\,\ref{fig:rgsphase}). 
While the spectrum  close  to minimum light can be well reproduced with
a three-temperature  plasma model, there are difficulties in finding a
suitable model that fits the high-resolution spectrum at maximum.
Surprisingly, the line ratios between the  CNO  elements are different
in different pulsation phases. Especially interesting is the 
variability of  the N\,{\sc vii}  Ly$\alpha$ resonance line.  This line
is much  stronger at the maximum than at the minimum of the light curve
(see Table\,\ref{tab:nvii}). We tried  to reproduce this line
variability by selecting models with different   temperatures or
additional absorption components, but could not find  a convincing
explanation for the observed line variability.  Allowing the nitrogen
abundance to be a free fitting parameter results  in a lower nitrogen
abundance during the phase  close to the minimum  ($4 \pm 1$ solar
nitrogen abundance in maximum versus $2 \pm 1$ in minimum).   This is a
puzzling result.

\begin{table}
\caption{Photon flux in the N\,{\sc vii} $\lambda 24.78$\,\AA\ line 
at different pulsation phase}
\begin{center}
\begin{tabular}{ll} \hline\hline
     Pulsation phase          & Photon flux [cm$^{-2}$s$^{-1}$] \\
\hline 
Average                 & $(2.6\pm 0.4)\times 10^{-5}$ \\
Maximum ($\phi=-0.9\,-\,0.2$)  & $(3.6\pm 0.7)\times 10^{-5}$ \\
Middle  ($\phi=0.75\,-\,0.9$) & $(2.3\pm 0.5)\times 10^{-5}$ \\
Minimum ($\phi=0.4\,-\,0.75$) & $(1.3\pm 0.6)\times 10^{-5}$ \\
\hline \hline    
\end{tabular}
\end{center}
\label{tab:nvii}
\end{table}

The location of X-ray emitting plasma can be constrained with the help of 
the lines from helium-like ions. These ions emit a group of three X-ray lines, 
consisting of a  forbidden ($f$), an intercombination ($i$), and a resonance ($r$) 
transition -- the  so-called $fir$ triplet.  The ratio of fluxes  $G=(f+i)/r$ is
sensitive to the temperature, while the ratio of fluxes between the
forbidden and the  intercombination component,  $R=f/i$, is sensitive to
the electron density and  the ultraviolet radiation field \cite{blum1972}: 
\begin{equation} 
R(r)=\frac{{\cal R}_0}{1+\phi(r)/\phi_{\rm c}+N_{\rm e}(r)/N_{\rm c}}. 
\label{eq:r} 
\end{equation} 
Here, $\phi$ denotes the
photo-excitation rate from the term 2s\,$^3$S to 2p$^3$P, and $N_{\rm
e}$ is the electron density. The constants ${\cal R}_0, \phi_{\rm c}$,
and $N_{\rm c}$ depend on atomic parameters and slightly on the
electron temperature \cite{blum1972,porq2001}. Most important, the
photo-excitation rate $\phi(r)$ scales with the mean intensity of the
radiation field at the wavelength  of the $f\rightarrow r$ transition, which
is typically in the UV. 

The $fir$ triplets of  Ne\,{\sc ix}, O\,{\sc vii}, and  N\,{\sc vi} are
present in the RGS spectrum of \xicma, with the O\,{\sc vii} triplet
being  the strongest one. Only in the O\,{\sc vii} triplet, the forbidden 
line is marginally detected, while forbidden lines in Ne\,{\sc ix} and 
N\,{\sc vi} were not detected.

We use the PoWR stellar atmosphere model to calculate the values of
$R(r)$ for Ne\,{\sc ix},  O\,{\sc vii} and N\,{\sc vi} as function of
the radial location in the   wind of \xicma. The photo-excitation  rates
$\phi(r)$ are computed at  each radius from the radiation intensity as
provided by the PoWR model,   which accounts not only for geometric
dilution, but  also for the  diffuse radiation field.
Figure\,\ref{fig:tau1} (right panel) shows for the O\,{\sc vii}  
triplet the predicted $R$ ratio  as function of the radial location of 
the emitting plasma. From the spectrum integrated over the  full
exposure time,  the best-fit value of the $R$ ratio indicates a plasma
location at  $\approx 6\,R_\ast$, while the 1$\sigma$ margins are
consistent with any   location between $1\,R_\ast$ and $10\,R_\ast$. 

\section{Discussion}

It is generally assumed that X-ray emission from hot massive stars 
originates from plasma heated by hydrodynamic shocks in a radiatively 
driven stellar wind. The radiation driven wind theory  predicts that 
the mass loss rate depends on the stellar parameters as 
\begin{equation}
\dot{M}\propto L_{\rm bol}^{\frac{1}{\alpha}}M_{\rm 
eff}^{\frac{\alpha-1}{\alpha}},
\label{eq:mdot}
\end{equation}
where $L_{\rm bol}$ is the stellar bolometric luminosity and 
$\alpha$ is a dimensionless number (which is about 2/3) representing 
the power-law exponent of the distribution function of line strength of 
the thousands of spectral lines driving the wind, 
and $M_{\rm eff} \approx M_\ast$ for B-type stars \cite{LC1999,kud1998}. 

During a pulsation cycle, the radius of \xicma\ changes by $4\%$ and its
effective temperature changes by $3.6\%$, with consequent changes for
the luminosity.  Recall that the luminosity is given by $L_{\rm
bol}=4\pi R_\ast^2 \sigma T_{\rm eff}^4$, where $\sigma$ is
Stefan--Boltzmann constant.  With $R_\ast$ and $T_{\rm eff}$ varying in
anti-phase, the bolometric luminosity changes by $\approx 7\%$ during a
pulsation cycle. According to equation\,(\ref{eq:mdot}), the mass-loss
rate may change by $10\%$, which is a similar amplitude as of the
observed X-ray pulsations. According to the wind theory, the wind 
velocity scales with the escape velocity, i.e.\ proportional to 
$\sqrt{M_{\rm eff}{R_\ast}^{-1}}$. Therefore it may also slightly change
during the pulsation.  The dynamical time scale for a stellar wind is
the flow time $R_\ast v_\infty^{-1}$.  For \xicma\ this time is about
2\,h. Hence, the wind is principally able to follow changes that are
triggered with the pulsation period (5\,h). From these considerations
one might, in principle,  expect that the stellar pulsations lead to 
periodic changes of the wind parameters and, consequently, of the X-ray
luminosity.  It is interesting to note that there is an established
empirical  correlation between bolometric and X-ray luminosity for OB
type stars,  $L_{\rm X}\sim 10^{-7} L_{\rm bol}$ \cite{pal1981}.  

However, while these scaling correlations are general, no X-ray
pulsations  from other stars were detected so far. For instance, a
dedicated study of the  X-ray light curve of $\beta$\,Cephei obtained
with \xmm\ did not reveal the presence of pulsations \cite{fav2009}. We,
therefore,  believe that the general scaling correlations   do not
explain the observed X-ray properties of \xicma.

The new \xmm\ observations of \xicma\ may provide important insights
into possible mechanisms of plasma heating. In massive stars that
possess  large scale magnetic fields, plasma heating is  commonly
explained by the magnetically confined wind shock mechanism  (MCWS)
\cite{bmb1997,ud2002}. This model predicts that the parameters of  X-ray
emission depend on mass-loss rate, wind velocity, stellar rotation,  
and especially on the magnetic  field strength \cite{mc2009,ud2009}. 

To check these predictions, we compared our data with RGS spectra of  
comparable quality that are available for two other magnetic
$\beta$\,Cep-type stars, $\beta$\,Cen (B1III) and $\beta$\,Cep (B2V).
$\beta$\,Centauri is a binary system consisting of two
$\beta$\,Cep-type  stars of nearly equal mass  \cite{aus2006}.  The
primary rotates significantly faster ($v\sin i =190\pm
20$\,km\,s$^{-1}$)  than the magnetic secondary  ($v\sin i =75\pm
15$\,km\,s$^{-1}$) \cite{aus2006, alecian2011}.  For its spectral 
class,  $\beta$\,Centauri has an average X-ray  luminosity, $L_{\rm
X}=1\times 10^{31}$\,erg\,s$^{-1}$. From the analysis of  its X-ray
spectrum, the temperature of the X-ray emitting plasma is $\left< kT
\right>\approx 0.3$\,keV, similar to the one measured in \xicma.  The
ratios of forbidden to  intercombination line strengths in He-like  ions
are also similar to those in \xicma\ \cite{raas2005}.

$\beta$\,Cephei is also a  binary system, which consists of a  magnetic
B1V  star as primary \cite{dbcep2001} and secondary of type  B6-8e 
\cite{schnerr2006}. The temperature of the X-ray emitting plasma   is 
also $\left< kT \right>\approx 0.3$\,keV, is similar to  \xicma\
\cite{fav2009}. The forbidden to intercombination line  ratio in He-like
ions is basically identical with those measured in \xicma.

The wind parameters are also quite similar in  $\beta$\,Cen,
$\beta$\,Cep and  \xicma. However the magnetic field strength of \xicma\
is at least one order  of magnitude higher than in other two objects.
Moreover, the viewing geometry  of \xicma\ may be special as we likely
look face on at the plane of the stellar  rotational equator. Yet, the
X-ray temperatures and the locations of  hot plasma are similar in all
three stars.

Besides MCWSs, another possibility for plasma heating in $\beta$\,Cep-type 
variables could be the deposition of mechanical energy from the stellar 
pulsations \cite{cas1996,Neilson2008}. 
%
However, in a study of a small sample of $\beta$\,Cep-type variables,  
no dependence of the X-ray luminosity on the pulsational period or 
amplitude was noticed \cite{osk2011}. 

The properties that make \xicma\  distinct from other $\beta$\,Cep-type 
variables are the supersonic speed of the pulsating photosphere, 
the high strength of the star's magnetic field, and the 
absence of non-radial modes in the stellar oscillations. We suggest 
that a successful explanatory model must take
these ingredients into account in order to understand the X-ray emission
and its periodic modulations observed in \xicma.

To summarize, we detect pulsations of the X-ray flux from \xicma\ in
phase with optical pulsations but with larger amplitude. The X-ray light
curves are similar in different energy bands. There are tentative
indications that changes of the X-ray emission are due to the plasma 
heating and cooling -- from the spectral analysis, somewhat higher 
temperatures are deduced at X-ray maximum as compared to X-ray minimum. 
The X-ray emitting plasma is located close to the photosphere, as
follows from  the cooling rate and from the analysis of line ratios in
He-like ions.

This first discovery of X-ray pulsations from a non-degenerate, massive
stars  will stimulate theoretical considerations for the physical
processes  operating in magnetized stellar winds. So far, the mechanism
by which the X-rays are affected by the  stellar pulsation remains
enigmatic.

\end{document}

%% file: bibdefinitions.tex
\def\araa{ARA\&A}%
\def\apj{ApJ}%
\def\apjl{ApJ}%
\def\apjs{ApJS}%
\def\aap{A\&A}%
\def\aapr{A\&A~Rev.}%
\def\aaps{A\&AS}%
\def\mnras{MNRAS}%
\def\pasp{PASP}%
\def\zap{ZAp}%